\documentclass{article}
\usepackage[T1]{fontenc}
\usepackage[utf8]{inputenc}
\usepackage{ismir} 
\usepackage{amsmath,cite,url}
\usepackage{amssymb}
\usepackage{graphicx}
\usepackage{color}
\usepackage{xurl}
\usepackage{arydshln}
\usepackage{subcaption}

\title{Finding the noise: Zero-shot AI Music Detection}

\oneauthor
  {Darius Afchar \hspace{10em} Romain Hennequin} {Deezer Research \\ \texttt{research@deezer.com}}

\newcommand{\ie}{\textit{i.\,e.,} }
\newcommand{\eg}{\textit{e.\,g.,} }

\newcommand{\cf}{\textit{cf.\,}}

\begin{document}

\maketitle

\begin{abstract}

We present a novel method for AI-generated music detection in scenarios where the models that generated the input samples are unknown to the detector (e.g., from a newly released service). Since 2023, there has been a multiplication of user-friendly AI-music generation services (e.g., Suno, Udio), along with regular updates and new features. There is thus a need to address synthetic content detection in an unsupervised way to adapt to this rapidly changing context. This angle has not been much studied in music yet.
We propose to study two tasks. First, discriminating between real and synthetic music. This may be approached in a one-class manner, namely, using some baseline real music and trying to determine what falls outside.
Second, zero-shot multi-class identification, which is more similar to an unsupervised clustering task on a mix of real and various AI-music generations, where the goal is to create coherent, high-purity clusters.
We propose a combination of a previously proposed artifact-extraction method, on top of which we apply non-negative matrix factorization and simple classification and clustering methods.
We achieve excellent performance on both tasks, showing that the proposed methods may be used to monitor large-scale catalogs that may receive AI-generated samples from various newly released generative models.

\end{abstract}

\section{Introduction}\label{sec:introduction}

Since the end of 2023, there has been a massification of AI-generated music with the launch of services such as \textit{Suno} and \textit{Udio}.
These services follow the paradigm of popular tech services like OpenAI's ChatGPT in that they propose to fully generate music in a matter of seconds from a text prompt, and are available online for free.
Since then, streaming services have seen a steady increase in the volume of synthetic content delivered\footnote{
We use the term synthetic as a short-hand for whether an AI service produced a music sample.
}.
For instance, Deezer revealed in June 2025 that 18\% of the music they received daily was AI-generated \cite{deezer_genAI}.
This number is now nearing 50\%, less than a year later \cite{deezer_genAI2}. The latter also reported that most of the listening of AI content was fraudulent. Many more AI music services have also emerged since then (e.g., \textit{Producers}, \textit{ElevenLabs}, \textit{Mureka}).

In this context, we need to explore AI music detection more.
The goal of this paper is not to discuss the philosophical and musical implications of whether AI music should or should not be treated similarly to real music.
Indeed, there are many ethical and legal challenges surrounding AI music services: \eg how copyright and royalties should work with AI content, how to deal with hybrids, whether using copyrighted music falls under "fair use" \cite{gebru2023artists, goetze2024ai, wei2024exploring, deezer_genAI3}.
Beyond these considerations, it seems critical to at least label AI content properly to inform listeners.
This aligns with the transparency regulations around AI \cite{AIActArticle50, alanoca2025comparing}, as well as broader discussions on the proliferation of \textit{"AI slop"}, across many fields beyond music (e.g., images, news, comments, fake videos).
We believe that research on detectors may help push forward the conversation about the scale of the where, what, and how AI music is published, which in turn, may help enlighten the aforementioned debates around generative content, the protection of artists, and the value of human creation \cite{klein2025provocations, pelly2025mood, cisac_genAI, sturm2025made}.

This topic remains relatively novel, and there is little prior work on AI-music detection specifically (\cf Section \ref{sec:related_work}).
To our knowledge, prior work has focused on supervised detection. This leaves several research gaps open.
For instance, how to deal with unknown music generation services?
This question is also relevant for detectors trained on data from specific services that have been updated since training.
For instance, the \textit{SONICS} dataset \cite{rahman2024sonics}, used in MIREX 2025\cite{mirex}, might already be considered deprecated, as it contains Suno \textit{v3.5} music samples, whereas the latter has recently released \textit{v5.5}.
It thus seems necessary to explore more zero-shot techniques.

In this paper, we present a novel method to achieve such an aim.
We define two tasks that we deem relevant to study in the context of AI-music detection. First, discriminating between real and synthetic music. This may be approached in a one-class manner: having some real music and trying to establish what falls outside.
Second, blind multi-class identification, which is more similar to an unsupervised clustering task on a mix of real and AI-music samples generated by various different models. 

In Section \ref{sec:related_work}, we present the related work. In Section \ref{sec:method}, we formulate the two tasks and propose to use a mix of generative artifact extraction and non-negative matrix factorization. In Section \ref{sec:exp}, we conduct our experiments on three datasets of real and synthetic music.
Our results are promising, achieving high accuracy on most datasets and tasks, while requiring almost no supervision.


\section{Related work}\label{sec:related_work}

Currently, there are relatively few work on the topic of AI-music detection \cite{zang2024singfake, afchar2025ai, afchar2025fourier, li2024audio, rahman2024sonics, cros2025ai, frohmann2025double, pascu2026echoes}. The topic is quite novel: the first papers were published in 2024, following the release of Suno at the end of 2023 which dramatically increased the volume of AI-generated music uploaded to the internet \cite{deezer_genAI, deezer_genAI2, aislop_404, aislop_guardian, aislop_time}. 
As mentioned, all the previous cited work regard supervised detection. To our knowledge, no paper has yet explored unsupervised settings.
Nevertheless, zero-shot learning and non-negative matrix factorization have been successfully employed in many other music and signal processing tasks (\eg \cite{bertin2007blind, wilson2008speech, fevotte2018single, parekh2024tackling}).

The motivations behind the development of AI detectors may be put into perspective through discussion on the  \textit{"AI Slop invasion"}. There exists a Wikipedia page that is quite exhaustive on that topic and listing many recent cases around generative AI in music \cite{enwiki:1350499922}: \eg the \textit{Velvet Sundown} and \textit{Breaking Rust} bands that topped the charts. Streaming service Deezer have also evidenced that a majority of streams on AI music found in their catalog turned out to be fraudulent \cite{deezer_genAI3}.

However, \textit{might all AI-music be slop?}
Several authors have dissected the notions of creativity and authorship in AI music with a more philosophical angle \cite{goetze2024ai, sturm2025made, pram2025opening, morreale2025reductive}, as well as the sociotechnical context around the provider of such services: \eg their tendency for extractivist data collection, reliance on low-paid annotation labor, and risk of job displacement for musicians \cite{morreale2021does, morreale2023data, morreale2024unwitting, morreale2026human, pelly2025mood, cisac_genAI}.

Beyond music, the impact of generative AI on society, politics, fake news, work and culture is actively discussed \cite{crawford2021atlas, snake_oil, wei2024exploring, klincewicz2025slopaganda, klein2025provocations, carbonell2025taylorisme, coeckelbergh2026technofascism, gautam2024melting, noroozian2025generative}.
Without opening too much the Pandora's box on the potential risks, harms, and opportunities around generative AI, we can at least note that there is global consensus on the need for greater transparency into synthetic generations. This principle been transcribed into European law in the \textit{Digital Service Act}, and later the \textit{AI Act}, as well as in several other countries \cite{AIActArticle50, alanoca2025comparing}.

Our contribution on AI music detection is motivated by this transparency goal. Our focus on unsupervised and zero-shot setting stems from the current lack of research in the existing literature.

\section{Method}\label{sec:method}

We provide some background on AI-music artifacts, formalize the tasks we propose to study, and present our method to achieve such an aim.

\subsection{Background}\label{sec:background}

\begin{figure}
    \centering
    \includegraphics[width=\linewidth]{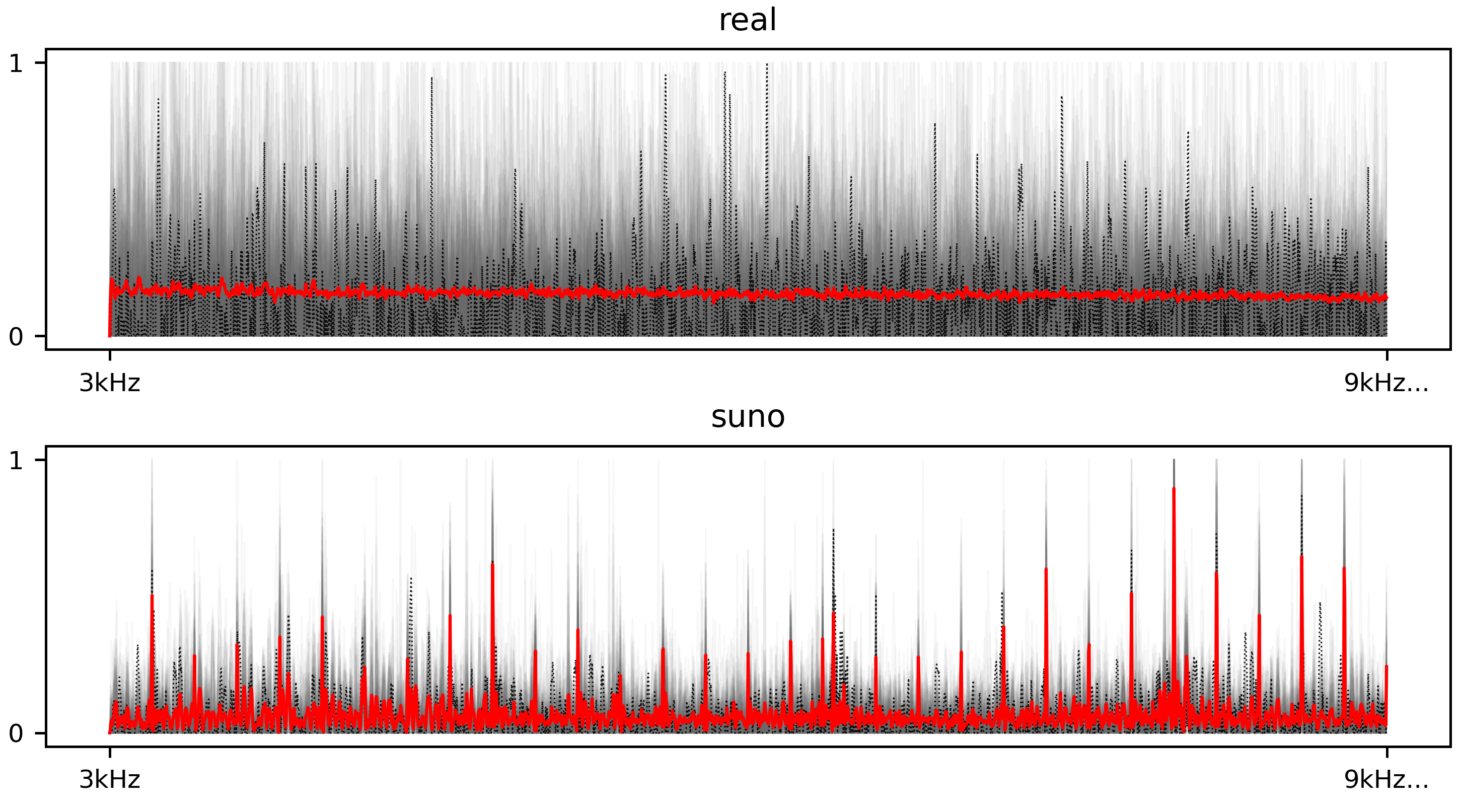}
    \caption{\textbf{Real and synthetic fakeprints.} We plot for each class a 100 random music fakeprints in transparent to see the distribution, one more sample in dotted black, and the average fakeprint in red. We zoom in on the frequency band 3kHz-9kHz for readability.}
    \label{fig:fakeprint}
    \vspace{-1em}
\end{figure}

To understand the reasonings behind our method, we must briefly introduce the AI-music artifact extraction process proposed in \cite{afchar2025fourier}.
Simply put, the deconvolution layers of convolutional neural networks were shown by the authors to leave small periodical energy peaks across all frequencies (\ie "checkerboard artifacts" \cite{odena2016deconvolution}). Their detection method thus relies on extracting such peaks profile (\eg see Figure \ref{fig:fakeprint}), and then learning a linear regression on top of it on a dataset of real and synthetic samples.

To obtain such profiles, a given music sample $x$ is processed in short-time Fourier transforms, and its power spectrogram is computed. The latter is temporally averaged to even out the melodic information. On this mean power curve, they subtract the local minima on sliding windows to extract the residual local peaks. The result for each music sample is a curve representing the time-averaged local variations in power per frequency.
A frequency cut is applied to only extract a band of interest, and finally, it is normalized by its max to result in a representation in $[0, 1]^d$. We refer to this processing as $\texttt{fakeprint}(x)$. The authors showed that this process was sufficient to extract the deconvolution artifacts in music tracks (if any are present) and obtain near-perfect detection scores, while remaining fast to compute and directly interpretable.

In this work, we propose to employ the $\texttt{fakeprint}()$ as a base representation for our zero-shot tasks.
In Figure \ref{fig:fakeprint}, we show the average fakeprints of Suno versus real music. As it may be seen, synthetic fakeprints exhibit the expected regular peaks, and in all of their samples. Meanwhile, real samples exhibit peaks---which are always created by the normalization step---but their localization seems uniformly scattered across frequencies between different samples. This results in an average fakeprint that is quite flat, denoting an absence of structure in real music's fakeprints.
This observed property is the key in our method to distinguishing synthetic content (consistent peak pattern) versus real content (random peaks placement).

\subsection{Tasks and assumptions}

In zero-shot learning tasks, the models do not use labels in their training \cite{chang2008importance, larochelle2008zero, palatucci2009zero}. Instead, some other \textit{a priori} information must be used on the data to separate them in a way that makes sense (\eg structural information, inductive biases, backbone pretrained on different data). Zero-shot may also mean learning on a subset of the labels, and trying to predict new unseen classes.
In this spirit, one-class classification describes a setting where samples from only one class are provided to the model \cite{de1998experimental, scholkopf2001estimating, noumir2012simple}. The goal is to learn the proper frontiers enclosing all given samples and to assign a binary label to all the remaining samples as being inside or outside of this support.  
In both cases, the model is blind to a part of or all of the input labels.

The \textit{a priori} knowledge we propose to use in this paper is that fakeprints may be a good representation of samples for the downstream task of blind synthetic content detection. This is informed by the previously reported good performance in supervised settings. We also test this assumption in Section \ref{sec:separability}.
We propose two variants of zero-shot tasks that we find relevant to study in our context of AI-music detection with regular new services and versions:

\vspace{1em}
\noindent\textbf{Task A.} We may first aim to distinguish real versus synthetic music in a binary way. We regroup all AI music under the target label "synthetic". One-class classification is very suited to this setting. Our goal is to learn the manifold of real music among a collection of samples that includes real music and several AI services music. We assume the class cardinality not to be too imbalanced. We also assume to be able to estimate a good upper bound to the number of all different classes that may be in a given dataset.

\vspace{1em}
\noindent\textbf{Task B.} To have more granularity, we may then aim to separate all classes altogether, namely all AI services, but this may also regard different versions of that service that may exhibit different artifact patterns. There, the real class is treated as one class among others. Zero-shot learning is very suited to this task, relying on \textit{a priori} information and structure analysis to cluster the music samples. Note that some labeled samples and metadata may later be used to interpret each cluster (if properly separated).

\vspace{1em}
As we next explain, both tasks may be tackled with two variants of the same method.

\subsection{Method for A (real versus synthetic)} \label{sec:method_a}

As hinted in the background section \ref{sec:background}, the key principle we rely on is that synthetic samples will follow some structured peak profiles (\eg roughly one per AI service) whereas real music sample will exhibit random peak localizations, akin to random noise.
Note that the "structures" we seek arise from statistical analyses, but not from the shape of the pattern themselves.

\vspace{1em}
\noindent\textbf{Dictionary learning.}
With this in mind, our idea is to employ a dictionary learning method that will extract patterns from the fakeprints. We denote as $X \in [0, 1]^{n\times d}$ a preprocessed dataset of $n$ fakeprints of length $d$ containing real and synthetic examples.
Given that fakeprints are positive, we propose to employ a \textit{non-negative matrix factorization} (NMF) \cite{lee2000algorithms, cichocki2009fast}. With the NMF, the goal is to approximate $X$ with a factorization $HW$. The matrix $H \in \mathbb{R}_+^{n \times f}$ is often referred to as containing \textit{activation} coefficients, while $W \in \mathbb{R}_+^{f \times d}$ contains the \textit{atoms} (or \textit{components}), namely the dictionary of patterns that are learned from the data. The factorization parameter $f$ is chosen as the upper bound for the total number of classes we expect to find in the data. In practice, we often have a reasonable estimate for $f$.

Even if $f$ is chosen too high, we add $\ell_1$ and $\ell_2$ regularizations (\ie an elastic net) in the NMF optimization:
\begin{equation*}
    \min_{H,W} \| X - HW \|_2^2 + \lambda_H \|H\|_{1,2} + \lambda_W \|W\|_{1,2}
\end{equation*}
with $\|.\|_{1,2} = \|.\|_1+\|.\|_2^2$. This enables to have a well-behaved dictionary of atoms. We only want an atom to be learned if a sufficient collection of fakeprints follow a given pattern. Conversely, we do not want each unique fakeprint of real music to be learned. Said differently, \textit{real music is treated as random noise} in the fakeprint representation. NMF has been used in audio with this denoising aim (\eg \cite{wilson2008speech}). 

After the denoising effect of the NMF, does this mean that synthetic samples will all have positive activations in $H$ and a corresponding component in $W$, whereas real samples have null activations? Almost. After some preliminary experiments, it turns out that some atoms are learned for real music: flat profiles without localized peaks. This makes sense as denoising results in approximating the mean envelope of real fakeprints. Therefore, we need one extra step to identify the fakeprints of real music.

\vspace{1em}
\noindent\textbf{Delocalized reconstruction error.}
Given the learned activations $H$ and atoms $W$, our idea is that $W$ will have learned some average peaks profiles: structured atoms with localized peak for synthetic samples, and flat envelopes to reconstruct the noise of real music (similar to Figure \ref{fig:fakeprint}).
We propose to destroy the learned peak structures in $W$ by convolving the atoms with a Gaussian kernel $G$. Since fakeprints represent mean power spectra, from a Fourier perspective, this Gaussian blur may be interpreted as a low-pass filter in cepstra. 
Then, we compare the reconstructions $\widetilde{X} = HW$ and $\widetilde{X}' = H(W*G)$, and define for all sample of index $i$, the error $r_i = \| \widetilde{X}_i - \widetilde{X}'_i \|_2$.
We expect $r_i$ to be low for real music and high for AI-music.
Our criterion $r$ is also very reminiscent of edge detection in computer vision. We can interpret it as finding whether the reconstructions $(\widetilde{X})_i$ have "edges", \ie peaks.

\vspace{1em}
\noindent\textbf{Threshold choice.}
So far, we have worked in a purely unsupervised manner. We now need to fix a threshold to distinguish whether $r_i$ correspond to a real or synthetic sample $i$.
We may use some examples of real samples to calibrate this. This is where this method turns into a one-class classification. Following previous literature \cite{scholkopf2001estimating}, given a small training collection of real examples, we take a quantile of a training set of errors $(r_i)$, for instance, 95\%. This allows to have a direct control of the false-positive rate in the detection of real music (\eg 5\%).

\subsection{Variant for B (multi-class clustering)}

In the second setting we study, we may also try to have more granularity in the patterns we have learned.
Coming back to our proposed method in section \ref{sec:method_a}, we have used a NMF to distinguish atoms with localized peaks (synthetic class) versus atoms without localized information (real class).
But a NMF can also be directly interpreted as learning a clustering. Therefore, we can already stop at this step to have a zero-shot clustering of the real and all the various possible AI-music classes.

Since the space learned in the NMF is latent and in high dimension, we propose to add a \textit{UMAP} to better visualize the learned representation \cite{mcinnes2018umap}. This technique is usually handy for separating clusters well, and when there is no need to have a meaningful metric on the cluster size and their relative distance. This method was also created to be easy to parametrize, with only one main parameter, which is suited for zero-shot learning.
Since we do not know in advance the number of clusters, we propose to employ the \textit{HDBSCAN} clustering \cite{campello2013density}, which is also quite straightforward to use with minimal required information on the data.
For task B, we have arguably proposed a much more straightforward combination of known techniques. We will see in the next section that this works well as is for AI-music identification.

Differently from the previous section, the variant we proposed for task B is purely zero-shot. We still have some main parameters to configure, which can often be estimated from the dataset size and the minimum expected cluster sizes.
We could use a few classes to tune these hyper-parameters and study the detection for the other unknown classes.
Afterwards, if the clustering works well, a few samples may also be employed to label the found clusters, identify the real music cluster, assign to each a known music service, or conversely identify a new cluster with an unknown artifact pattern (\ie few-shot learning). We leave these options as future work and use reasonable default parameters in this paper.

\section{Experiments}\label{sec:exp}

We present our datasets, some preliminary supervised results to check if the data is separable using the fakeprints, and the experiments for our two tasks and method.

\subsection{Datasets}\label{sec:dataset}

As exposed in Section \ref{sec:related_work}, there are relatively few work published on AI music detection, which also means few published datasets (\cf the survey \cite{li2024detecting}).
For our tasks, we first consider the dataset from \cite{afchar2025ai}, which provided autoencoded samples of the \textit{Free Music Archive} (FMA) \cite{fma_dataset} through 4 models: \textit{"Encodec, DAC, Musica, GrifMel"}.
In the original dataset, the autoencoded samples are available in several compression settings. We only pick the highest quality for each class. We use the "small" split of \textit{FMA} with 8000 samples per class.
We refer to it as \textit{FMA-AE}.

The previous dataset does not contain any sample from popular AI services.
We had first considered using the \textit{SONICS} dataset \cite{rahman2024sonics}---which was proposed in last's year \textit{MIREX} challenge. However, it only contains two classes (Suno and Udio). Even more critical, all audios in SONICS have been resampled to 16kHz, which results in a unsuited cutoff in the frequency bands of interest for fakeprints. Alternatively, we found a recent dataset in a publication still in preprint: \textit{Echoes} \cite{pascu2026echoes}. This dataset contains 10 classes from online services: \textit{"DiffRhythm, Suno, Brev, ACEStep, Producer, Udio, SongGen, AudioLDM, Mubert, and StableAudio"}.
Despite its relatively small size (3577 tracks in total), we found it interesting for our purpose to test our method on a variety of different services.
Since this dataset does not provide real audio tracks, we resort to including samples from the \textit{FMA}, as was previously done in the literature \cite{afchar2025fourier}. All music tracks are resampled to 44.1kHz.
 
Finally, for completeness, we create a dataset including the two most popular services Suno and Udio, as well as four more online services that had not been collected in Echoes yet quite popular: \textit{Lyria 3}, \textit{ElevenLabs}, \textit{Mureka} and \textit{Riffusion}.
We fetch a few thousand tracks for each of these services, as well as some metadata (\eg sub-versions). We resample them all to 44.1kHz. We dub this dataset \textit{PopularAISet}\footnote{\label{foot:repo}Will be made public after acceptance.}.

\subsection{Settings}\label{sec:settings}

We use the similar parameters to configure the fakeprints as in the original paper (\eg a \texttt{n\_fft} of $2^{14}$ for the STFT), and a frequency band of [3kHz, 15kHz].
This results in fakeprint vectors of length 4458.

In Task A, the Gaussian kernel is picked to be "quite large" relative to the single peaks we are trying to smooth, namely a radius 10 in our experiments. The number of NMF components are set to 20 since we know that this was a proper upper bound of the number of classes in our considered datasets.
In Task B, the \texttt{n\_neighbors} parameter of the UMAP is arbitrarily fixed at 20. And the minimum cluster size of the HDBSCAN is fixed at $n/5f$, an arbitrary fraction of the average expected cluster size ($n/f$).

We use the same set of settings for all three datasets.
The rest of the chosen parameters may be found on our code repository\textsuperscript{\ref{foot:repo}}.

\subsection{Preliminary separability test}\label{sec:separability}

Before moving to our two unsupervised tasks, we wanted to verify this paper base assumption that the fakeprint representation was suitable and could linearly separate AI music tracks from real music tracks with the datasets we study.
Compared to \cite{afchar2025fourier}, we test 18 different AI models in total, which includes many that the latter method has not been evaluated on.
This preliminary section thus serve as a quick reproducibility check of past work, as well as a quick overview on the extension of detection scores to models untested in the literature yet.

Following \cite{afchar2025fourier}, we train logistic regressors on fakeprints computed on the tracks of our datasets. Note that we operate in a supervised binary setting, not a multiclass one. We take the real class of each dataset and train a regressor for each available synthetic class.
The selected data is split in a 80\%-20\% fashion, and we compute the \textit{equal error rate} (EER) on the predictions.

The results are provided in Table \ref{tab:prelim}.
First, we insist on the fact that the Echoes dataset is very small. Therefore, after the train-test split, there only remain between 30 and 60 samples in each test sets. The displayed scores for Echoes should be taken with a grain of salt and are to be read as a quick sanity check.
Second, overall, the results in the supervised detection setup look very good, and comparable to previous work. This is reassuring that the fakeprint does what we want to achieve.
Finally, some specific classes have significantly worse linear separation scores that the others: Mubert in \textit{Echoes}, and Mureka in the \textit{PopularAISet}. We can anticipate those not to be well detected in the next unsupervised settings.

\begin{table}

    \small

    \centering

    \begin{minipage}[t]{0.48\linewidth}
        \vspace{0pt}
        \centering
        \begin{tabular}{|l|c|}
            \hline
            Dataset / class & EER ($\downarrow$) \\
            \hline
            \hline
            \textbf{FMA-AE} & \\
            Encodec & 0.1\% \\
            Musika & 0.0\% \\
            DAC & 0.0\% \\
            GrifMel & 1.0\% \\
            \hline
            \hline
            \textbf{PopularAISet} & \\
            Suno (\textit{Pop.AISet}) & 1.3\% \\
            Udio (\textit{Pop.AISet}) & 0.2\% \\
            Lyria 3 & 0.3\% \\
            Riffusion & 0.2\% \\
            ElevenLabs & 2.0\% \\
            Mureka & 8.7\% \\
            \hline
        \end{tabular}
    \end{minipage}
    \hfill
    \begin{minipage}[t]{0.48\linewidth}
        \vspace{0pt}
        \centering
        \begin{tabular}{|l|c|}
            \hline
            Dataset / class & EER ($\downarrow$) \\
            \hline
            \hline
            \textbf{Echoes} & \\
            Suno (\textit{Echoes}) & 0\% \\
            Udio (\textit{Echoes}) & 0\% \\
            Brev & 0\% \\
            StableAudio & 0\% \\
            ACE-Step & 0\% \\
            AudioLDM & 0\% \\
            DiffRhythm & 0\% \\
            SongGen & 0\% \\
            Producer & 0\% \\
            Mubert & 30\% \\
            \hline
        \end{tabular}
    \end{minipage}

    \caption{\textbf{Preliminary supervised binary detection scores.} The scores for Echoes are written with a minimal precision due to a lack of samples in the test sets.}

    \label{tab:prelim}

    \vspace{-1em}

\end{table}

\subsection{Results}\label{sec:results}

We present our results on the two tasks, for our method and its variant. As reminder, Task A is binary and one-class, while Task B is multi-class and zero-shot.
Note that we do not have any baseline model to compare our work to since this unsupervised angle is novel.

\vspace{1em}
\noindent\textbf{Task A: real versus synthetic.}
The only hyperparameter in our method A is the threshold on $r_i$, which can be interpreted as a target false positive rate (FPR) to bound the observed distribution of the real class.
In these experiments, we display the accuracy of detection for the synthetic classes, while targeting an FPR of 10\% and 1\%.
To achieve this, we compute these quantiles on half of the real examples and use the rest as test. As a reminder, we did not use any of the synthetic data.
We only discriminate the typical statistical patterns of real music from those that fall outside a defined boundary and may not be real music.

The results are presented in Table \ref{tab:oneclass}.
As a first observation, without any information on the synthetic class, it is hard to anticipate whether a target FPR of 10\% or 1\% is "going too far" and catching synthetic samples in the real class support: it worked in both cases on \textit{FMA-AE}, but degraded the performance for an FPR at 1\% for the other two datasets. 
The aggregated scores for the synthetic classes do not mean much as is, and the performances are better explained by looking at the breakdown for each model: Overall, the results are very good for all AI services, but for the two models \textit{Mubert} and \textit{Mureka} that we had already flagged in the linear separability test. Unsurpringly, the detection has worsened compared to Table~\ref{tab:prelim}: when the NMF denoised the fakeprints, the already poorly separated synthetic fakeprints were assigned to the flat atom\footnote{...which we could confirm by inspecting the learned activations $H$.} and thus assigned the real label.
Conversely, many results turned out to be as good as in the supervised setting. For instance, for Suno in the \textit{PopularAISet}, an EER of $1.3\%$ seems to match the $98.5\%$ for real versus $99.8\%$ for Suno.

All in all, these experiments are quite satisfying given the relative simplicity of our proposed criterion and method.
A 10\% FPR threshold may be considered large for a real-world application (\eg analyzing a streaming music service catalog). However, the main motivation of this work is to provide a prospective tool to be used as a complement to a given supervised detection model to help identify new AI services that would evade its radars, not to replace it. For instance, this may be used to raise an alarm on a swarm of undetected synthetic music all coming from the same record label or artist.

\begin{table}
    \small
    \centering
    \begin{tabular}{|l|c|c|}
        \hline
        Dataset and class & Acc @10\% & Acc @1\% ($\uparrow$) \\
        \hline
        \hline
        \textbf{FMA-AE} & & \\
         Real & 90.0\% & 99.4\% \\
         Synthetic & 99.4\% & 98.1\% \\
         \hdashline
        $\hookrightarrow$ Encodec & 99.9\% & 99.8\% \\
        $\hookrightarrow$ Musika & 100\% & 100\% \\
        $\hookrightarrow$ DAC & 99.9\% & 99.2\% \\
        $\hookrightarrow$ GrifMel & 97.8\% & 93.5\% \\
        \hline
        \hline
        \textbf{Echoes} & & \\
         Real & 90\% & 100 \% \\
         Synthetic & 93\% & 53 \% \\
         \hdashline
        $\hookrightarrow$ Suno (\textit{Echoes}) & 100\% & 67\% \\
         $\hookrightarrow$ Udio (\textit{Echoes}) & 87\% & 3\% \\
         $\hookrightarrow$ Brev & 100\% & 58\% \\
         $\hookrightarrow$ StableAudio & 98\% & 4\% \\
         $\hookrightarrow$ ACE-Step & 100\% & 95\% \\
         $\hookrightarrow$ AudioLDM & 100\% & 83\% \\
         $\hookrightarrow$ DiffRhythm & 100\% & 90\% \\
         $\hookrightarrow$ SongGen & 100\% & 46\% \\
         $\hookrightarrow$ Producer & 100\% & 39\% \\
         $\hookrightarrow$ Mubert & 5\% & 0\% \\
         \hline
         \hline
         \textbf{PopularAISet} & & \\
         Real & 90.4\% & 98.5\% \\
         Synthetic & 91.6\% & 80.4\% \\
         \hdashline
         $\hookrightarrow$ Suno (\textit{PopularAISet}) & 99.9\% & 99.8\% \\
         $\hookrightarrow$ Udio (\textit{PopularAISet}) & 99.4\% & 96.4\% \\
         $\hookrightarrow$ Lyria 3 & 98.8\% & 75.6\% \\
         $\hookrightarrow$ Riffusion & 99.9\% & 99.6\% \\
         $\hookrightarrow$ ElevenLabs & 93.3\% & 77.5\% \\
         $\hookrightarrow$ Mureka & 48.5\% & 14.9\% \\
         \hline
    \end{tabular}
    \caption{\textbf{One-class detection accuracy scores} on the three datasets at the chosen FPR 10\% and 1\%. For each dataset, we highlight the aggregate scores as well as the breakdown per each AI generator model.}
    \label{tab:oneclass}
    \vspace{-1em}
\end{table}

\begin{figure}
    \centering
    \begin{subfigure}[t]{\linewidth}
        \centering
        \includegraphics[width=0.85\linewidth]{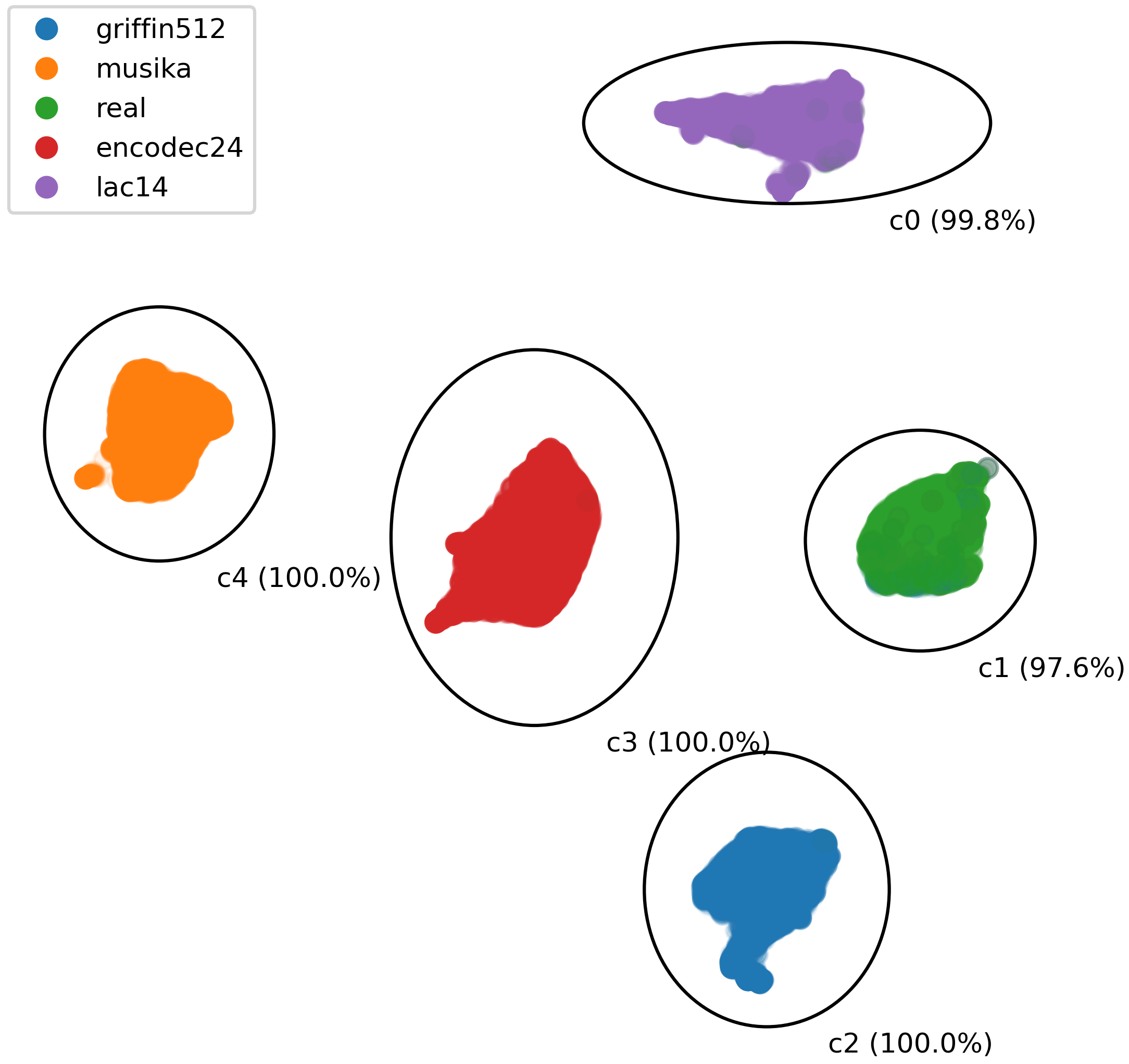}
        \caption{\textit{FMA-AE}.}
    \end{subfigure}

    \vspace{2em}
    \begin{subfigure}[t]{\linewidth}
        \centering
        \includegraphics[width=\linewidth]{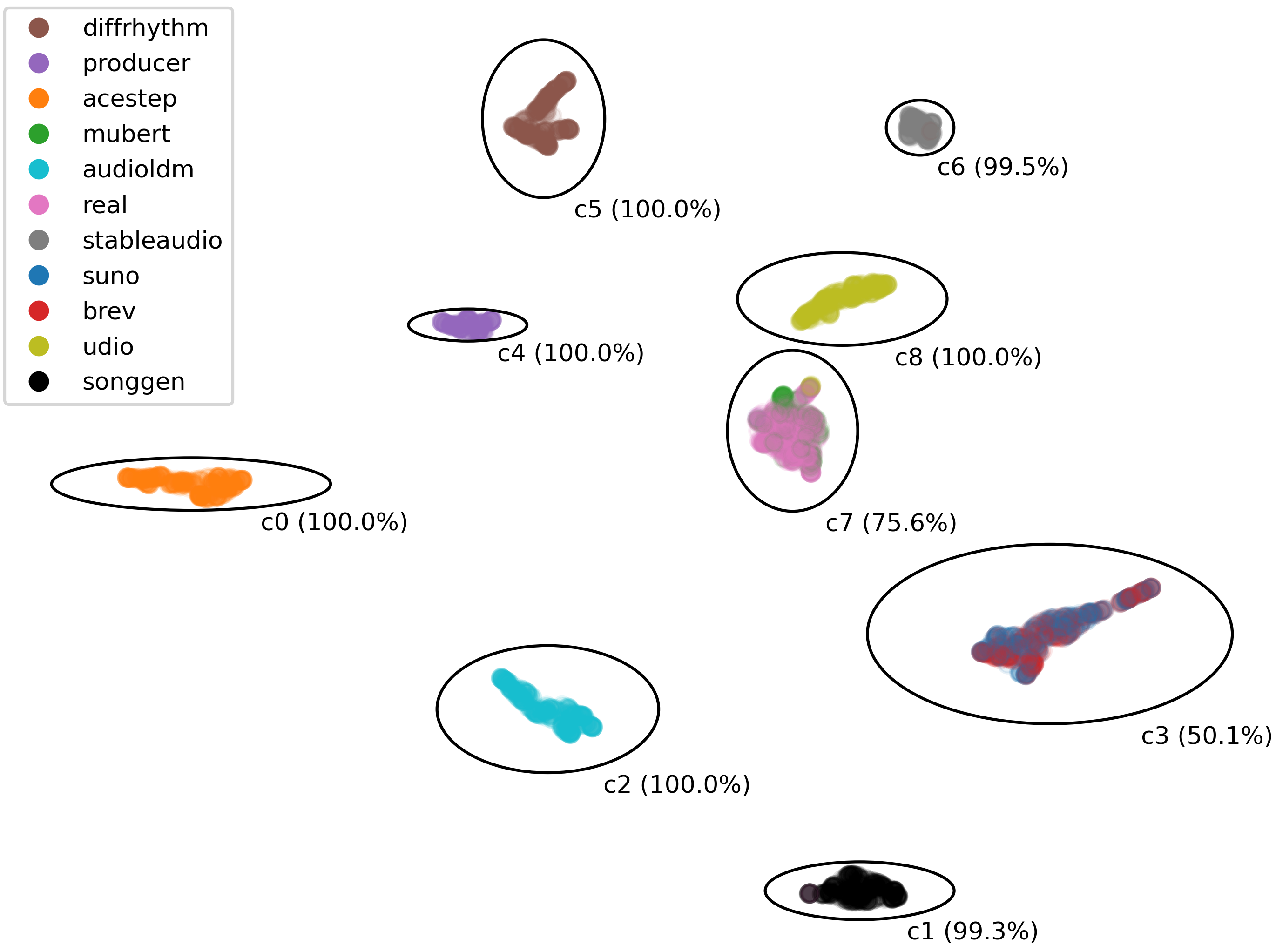}
        \caption{\textit{Echoes}.}
    \end{subfigure}

    \vspace{2em}
    \begin{subfigure}[t]{\linewidth}
        \centering
        \includegraphics[width=\linewidth]{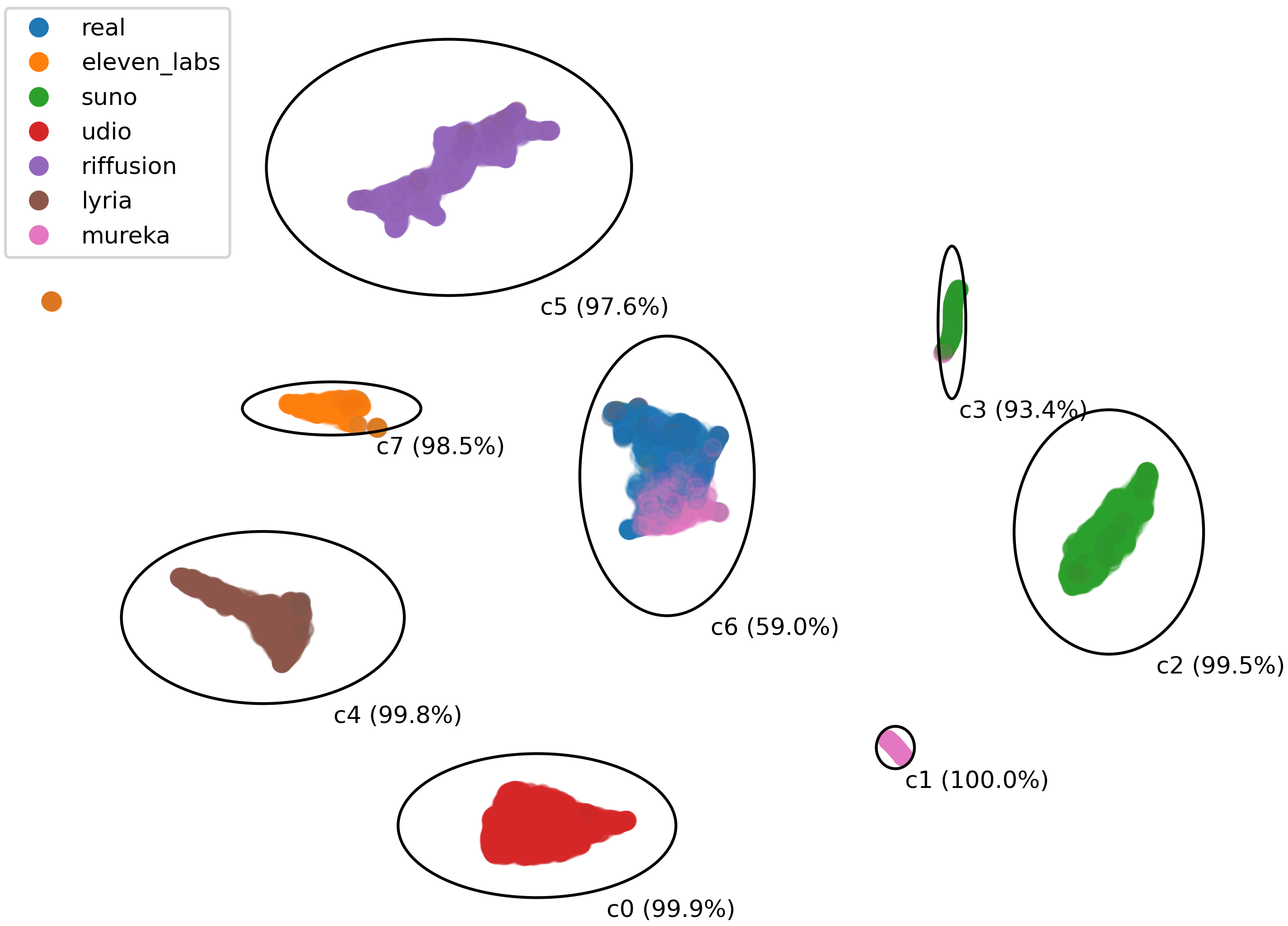}
        \caption{\textit{PopularAISet}.}
    \end{subfigure}
    
    
    \caption{\textbf{Clustering results for task B.} The colors indicate the ground-truth classes for each point. The ellipses are drawn around each cluster computed by the HDBSCAN. A cluster index "cX" is displayed for reference, as well as a purity score for each cluster. }
    \label{fig:clustering}
\end{figure}

\vspace{1.5em}
\noindent\textbf{Task B: multiclass clustering.} We now move to the task B and the variant on the method that involves clustering.
We find that the best way to present the result is to display the computed clusterings in Figure~\ref{fig:clustering}. We indicate the purity ratio in each cluster (\ie the ratio of the major label in a cluster) as an indication of consistency.

On the \textit{FMA-AE} dataset, the five classes are almost perfectly separated, with some punctual outliers (indicated with the less than 100\% purity).

On the \textit{Echoes} dataset, almost all classes are well separated, except for the cluster \texttt{c7} of mixed class \textit{real}+\textit{Mubert}, as well as another mixed cluster \texttt{c3} of \textit{Suno}+\textit{Brev}. The 75.6\% purity in the real class (instead of half), is due to the class size imbalance of the dataset (\ie 3 times more real samples than Mubert ones). The confounding on Mubert is consistent with our previous experiments, which showed that fakeprints from Mubert were not linearly separable from those of the real class.
We were curious about the mix of Brev and Suno. After searching online, we found out that \textit{"Brev.ai [...] uses Suno V3.5 technology to create original music from text descriptions."}\footnote{\url{brev.ai/ai-song-generator}}
Our aim was to potentially detect sub-versions of models to get a better granularity than the ground-truth labels. It turns out our method may also be used to identify services that rely on the same generator, which was an unanticipated but interesting side-effect.

On the third \textit{PopularAISet}, almost all models are properly clustered. Interestingly here, \textit{Suno} samples are split into two clusters \texttt{c3} and \texttt{c2}. After manual checking, all samples in \texttt{c3} correspond to \textit{Suno-v3.5}, while \texttt{c2} contains \textit{Suno-v4.5} and \textit{Suno-v5} samples. This suggests that the artifact patterns are different between versions, which seems to indicate a change in model architecture.
As anticipated from the separability tests, the cluster \texttt{c6} contains a mix of \textit{real}+\textit{Mureka} samples. Surprisingly, one subset of Mureka samples are clustered away in \texttt{c1}. We manually inspected the files and found out that all these were generated with the latest version \textit{V9} of Mureka (differently from the other dataset samples generated with \textit{V7.6}, \textit{V8} and \textit{O2}).
We wonder if the fact that it is the latest version of this service that includes detectable artifacts in the fakeprints could be linked to the use of a watermarking technology by the service to flag their generations, or if this is purely linked to architectural changes.


All in all, our proposed approach shows good properties on these three datasets. As a reminder, all of this was done in a purely zero-shot fashion. The metadata and version information from the \textit{PopularAISet} were only checked afterward to interpret the resulting clustering.

\section{Conclusion}\label{sec:conclusion}

We show that previous work conducted on AI music detection can be well extended to the zero-shot territory. Our method is fast, does not use deep learning, and can serve as a complement to existing supervised detectors to track newly released versions and models, paving the way for the exhaustive monitoring of AI-generated music in deliveries to large-scale catalogs.
One limitation of our work is the need for a sufficiently large sample collection for a synthetic class to detect it (otherwise, it is treated as noise and thus as real). In future work, we may want to investigate the shape in the fakeprints themselves (\eg detecting periodicities).
We should also further investigate why samples from Mureka (before \textit{v9}) and Mubert do not seem to be detected by the fakeprints and improve this technique.
Other future work may include addressing robustness to hybrid content, an open question in this field.

\clearpage

\clearpage

\bibliography{ref}

\end{document}